# An optical interferometer with wavelength dispersion

T.R. Bedding[1,2], J.G. Robertson[1], and R.G. Marson[1,3]

[1] School of Physics, University of Sydney 2006, Australia
[2] European Southern Observatory, Karl-Schwarzschild-Str. 2, D-85748 Garching bei München, Federal Republic of Germany
[3] CSIRO Division of Radiophysics, PO Box 76, Epping 2121, Australia

**Abstract.** MAPPIT is an optical interferometer installed at the coudé focus of the 3.9 m Anglo-Australian Telescope. The instrument combines non-redundant masking with wavelength dispersion and is able to record fringes simultaneously over a wide bandwidth. For typical observations centred near 600 nm, the bandwidth is $\Delta\lambda = 55$ nm and the spectral resolution is $\delta\lambda = 3$ nm. This paper describes the instrument and the data processing methods and presents some results. We find the star $\sigma$ Sgr to be a close binary; the system is only partially resolved, with a separation of $(11.5 \pm 2)$ milliarcsec (assuming the components to have equal magnitudes). We also give angular diameter measurements of two red giant stars, $\alpha$ Sco and $\beta$ Gru. The observations of $\beta$ Gru (spectral type M5 III) resolve the star for the first time and give an equivalent uniform-disk diameter of $(27 \pm 3)$ milliarcsec.

## 1. Introduction

Optical interferometry can be used to overcome the limits imposed by atmospheric turbulence and so allow large telescopes to achieve diffraction-limited resolution. To do so, one must compensate for perturbations in the wavefront of the light which result from its passage through the atmosphere. These perturbations arise because the refractive index of the atmosphere is continually fluctuating, primarily due to turbulent mixing of regions of air with differing temperatures.

The most direct way to compensate for wavefront perturbations is with adaptive optics. This technique uses a deformable mirror to make real-time corrections, with some fraction of the light being diverted to monitor the continually changing shape of the wavefront. Although adaptive optics shows great promise for improving the angular resolution of telescopes beyond the seeing limit in the infrared, its application to visible wavelengths is more difficult because of the large number of actuators required (see Beckers 1993 for a recent review).

Passive methods for attaining high angular resolution rely on recording the distorted image in a succession of short exposures and processing them off-line. This technique is known as speckle interferometry (Labeyrie 1987) or, if closure phases are also retrieved, as speckle masking (Weigelt 1991). Recently, it has been suggested that there are advantages in modifying the telescope pupil, at least for some observations. Non-redundant masking (Haniff et al. 1987; Nakajima et al. 1989) involves recording the short-exposure interferograms through a pupil mask containing a small number of holes, arranged so that all the baseline vectors are distinct. The interferograms are analysed to determine the power spectrum and closure phases of the object, which can be used to reconstruct a true diffraction-limited image. The advantages and disadvantages of aperture masking have been reviewed by Haniff (1994; see also Bedding et al. 1993). Here we merely summarize the main points.

One advantage of using a non-redundant aperture mask is that, for bright objects, it increases the signal-to-noise ratios of the power spectrum and closure phase measurements relative to observations with an unobstructed aperture. This is despite the fact that much of the light is blocked by the mask. Another advantage is that it improves the accuracy with which one can correct for variations in atmospheric seeing, something which is often the limiting factor in high-resolution imaging. This feature arises because the visibility of fringes obtained through a mask with small holes is relatively unaffected by changes in the seeing-cell size. This contrasts with case of a large aperture, where the transfer function is much more sensitive to fluctuations in the seeing (Buscher 1988).

The main drawbacks of non-redundant masking relative to a fully-filled aperture are the restriction to bright objects and a less efficient coverage of spatial frequencies. However, for simple objects such as multiple and barely-resolved stars, adequate spatial frequency coverage can be obtained by combining observations made with different masks and/or with the masks rotated to several different position angles on the sky. Other authors have discussed using a partially-redundant mask such as a thin slit (Aime & Roddier 1977; Buscher & Haniff 1993) or an annulus (Haniff & Buscher 1992).

Experiments in aperture masking with the Anglo-Australian Telescope were begun in the mid 1980s (Frater et al. 1986). This has culminated in the construction of MAPPIT (Masked APerture-Plane Interference Telescope), which we have used to investigate interferometry with non-redundant masks and also with slits. An important feature of MAPPIT is the use of a prism to disperse the interference pattern in wavelength. This allows us to overcome a restriction common to all forms of interferometry, namely the requirement that one observe over a narrow band of wavelengths. This paper describes the instrument and processing methods and presents some observational results.

## 2. The instrument

MAPPIT is situated in the west coudé room of the 3.9 m Anglo-Australian Telescope (AAT). Operating at the coudé focus has the drawback of light loss from reflections off the three extra flats in the mirror train. Nevertheless, the advantages of working in the coudé room are great. The components of MAPPIT are mounted on two optical rails which are supported by a heavy steel frame attached to the telescope foundation, providing excellent vibrational isolation. Using optical rails gives great versatility, since the optical configuration can be easily changed. In addition, the convenience of having full access to the instrument before and during observations has provided much flexibility and scope for experiment. Other workers conducting aperture-masking experiments at coudé have noted a loss in fringe visibility which they suggest may be due to the long light path (Kulkarni 1988). We have not found MAPPIT observations to be affected by this problem.

The main components of MAPPIT are shown in Fig. 1, while Fig. 2 shows the front part of the instrument. We now proceed to describe the individual subsystems in some detail.

### 2.1. Pre-focus optics

The atmospheric dispersion corrector (ADC; not shown in Fig. 2) consists of a pair of zero-deviation Risley prisms which are independently rotatable using stepping motors under microprocessor control. Their positions are chosen so as to introduce a dispersion which approximately cancels that of the atmosphere (Wynne & Worswick 1990).

Because MAPPIT uses wavelength dispersion, the aperture mask consists of a fixed one-dimensional array of holes. We therefore use a dove prism as a field rotator to obtain measurements at different position angles on the sky. The dove prism has a square cross-section ($20 \times 20$ mm) and is mounted on a rotation stage $\sim 300$ mm in front of the coudé focus. Such a prism should generally be used in collimated light to avoid an astigmatic image. In MAPPIT this is not practicable because the collimated beam has a diameter of 25 mm, which would require a large and expensive prism. The small amount of astigmatism introduced by placing the dove prism in the converging beam has not proved to be a problem.

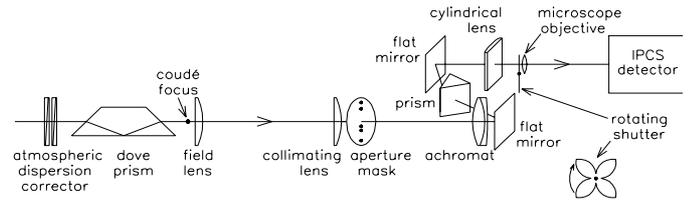

**Fig. 1.** Schematic view of the main components of MAPPIT (not to scale)

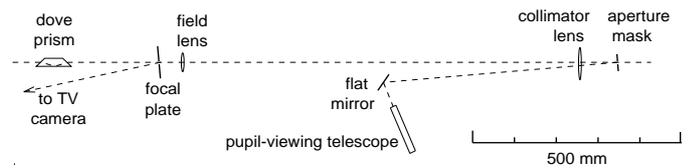

**Fig. 2.** Plan view of the front end of MAPPIT showing the location of the dove prism, TV guiding system and the optics for re-imaging and viewing the AAT pupil (approximately to scale). The focal plate used for TV guiding is at the coudé focus. The ADC (not shown) lies 800 mm to the left of the dove prism.

For acquisition and guiding we use a high-gain television camera focused on a polished stainless steel plate at the coudé focus. The focal plate has a central aperture of diameter 3 mm ($\sim 4''$ on the sky) and, with the TV gain turned up and the focal plate tilted to reflect starlight towards the camera, the wings of the seeing disk are easily visible as a ring on the TV monitor to use for guiding. For alignment and testing of MAPPIT, it is very useful to be able to generate an artificial star. To do this, we replace the focal plate with a pinhole and illuminate it with a tungsten-filament lamp. A 100 W lamp with a pinhole of diameter 10–50 $\mu$m produces excellent fringes. In addition, replacing the white-light source by a copper-argon arc lamp allows us to obtain spectra for wavelength calibration.

### 2.2. Re-imaging the pupil

The combination of field and collimator lenses serves to form an image in collimated light of the telescope pupil on the aperture mask, as shown in Fig. 3. When observing a bright star, the image of the AAT pupil is easily visible on the aperture mask. As indicated in Fig. 3, one can also see the shadows of the vanes which support the secondary mirror structure.

It is very important during observations to ensure that all the mask holes lie within the annulus of the primary mirror and none is obscured by the vanes. We achieve this by moving the field lens. The mount holding this lens has adjustments for the lateral and vertical directions which allow one to translate the pupil image with respect to the aperture mask. The field lens is placed close to the coudé focus so that this adjustment has only a small effect on the position of the star's image on the detector. Ideally, the field lens would be exactly at the focus; however,

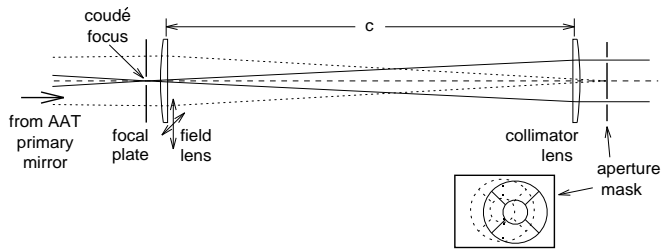

**Fig. 3.** Side view of the optics used to re-image the AAT pupil. The solid lines show rays from a star that is being imaged at the coudé focus; the rays become parallel after passing through the collimator lens. The dotted lines show rays from the centre of the primary mirror converging to form a pupil image on the aperture mask. Translating the field lens, as indicated by the double arrows, moves the pupil image with respect to the mask. The diagram is to scale in the horizontal direction; the vertical scale has been exaggerated for clarity.

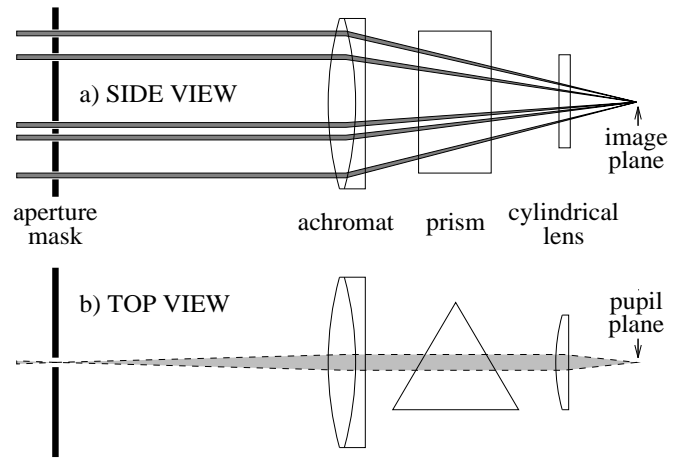

**Fig. 4.** Optics used to produce dispersed fringes, not to scale: **a** side view (the interference direction), **b** top view (the dispersion direction). For simplicity, two flat mirrors are not shown and the prism is depicted as having zero deviation (see Fig. 1 for a three-dimensional view).

that position is occupied by the focal plate used for TV guiding (see above).

The alignment of the pupil with respect to the aperture mask must be checked every time the position angle of the dove prism is changed or the telescope moves to a different star. It is therefore essential to have a convenient method of viewing the pupil image, especially for ensuring that none of the holes are obscured by vanes. The system for monitoring the pupil image is shown in Fig. 2. It consists of a small telescope that views the aperture mask via a flat mirror. The aperture masks are made of stainless steel and have a high degree of specular reflectivity so that by tilting the mask by $\sim 5°$, most of the reflected light can be directed into the viewing telescope.

Behind the mask we have mounted a photo-luminescent tile which glows when connected to a voltage. When aligning the pupil before an exposure, it is a simple matter to apply the voltage and swing the tile into the beam. It can then be seen through the holes in the mask, making it easy to identify the holes and position them relative to the pupil. This system has proved very satisfactory for all the stars we have observed to date.

### 2.3. The aperture mask

Most of our observations have used a pupil mask with five apertures with diameters of $\sim 0.4$ mm, which projects to $\sim 5$ cm on the primary mirror. The apertures are arranged in a linear non-redundant array, with the spacings chosen to optimize the spatial frequency coverage (Marson 1994). The masks themselves are made commercially from stainless steel using a photochemical etching process, from artwork produced by a laser printer. The accuracy achieved is necessary if the fringes are to be dispersed in wavelength. The process also allows us to generate square holes, which are better suited to the dispersion system than circular ones (see below).

### 2.4. The wavelength-dispersion system

Combining optical interferometry with wavelength dispersion has been used successfully in two-element long-baseline interferometers, both at the I2T (Thom et al. 1986) and the GI2T (Mourard et al. 1989). In the laboratory, a five-element non-redundant interferometer with wavelength dispersion has been developed by Cruzalèbes et al. (1993). Experiments using dispersion with single-telescope interferometry (speckle spectroscopy) have been carried out by Grieger & Weigelt (1992).

Dispersed fringes in MAPPIT are produced using a novel combination of image-plane and pupil-plane imaging. Figure 4 shows the relevant optics, from the aperture mask to the input focus of the microscope objective. Dispersed fringes recorded using a pinhole source as an artificial 'star' are shown in Fig. 5.

The operation of the system can be analysed by considering the interference and dispersion directions separately. The optics in the interference direction form a conventional image-plane interferometer (see Fig. 4a). The beam is already collimated and the achromat produces an image of the star crossed by interference fringes. In the orthogonal direction we have the spectrograph, but it is a dispersed *pupil* image that is formed by the optics. As shown in Fig. 4(b), the role of the spectrograph slit is played by the mask holes, seen 'end on.' The achromat now becomes the spectrograph collimator, with a cylindrical lens serving as the spectrograph's camera lens.

A major advantage of this optical configuration is the stability of the wavelength scale on the detector. This contrasts with using a dispersed image of the star itself, where the spectrum shifts continually with fluctuations in the seeing. Another benefit of the system is that the amount of anamorphic magnification required is substantially less than in other configurations. Anamorphic magnification in a dispersed system is necessary to compress the image in the dispersion direction and so accommodate a reasonable number of spectral channels on the detector. We are restricted to using singlet cylindrical lenses because of cost (cylindrical achromats must be custom-made), so it is desirable to use low magnifications to minimize aberrations. This is possible with the MAPPIT design because the spectrograph slit, consisting of the projected array of mask holes, has a width of $\sim 0.4$ mm. This is already less than one third the size of the stellar seeing disk at coudé, which means that less anamorphic magnification is required.

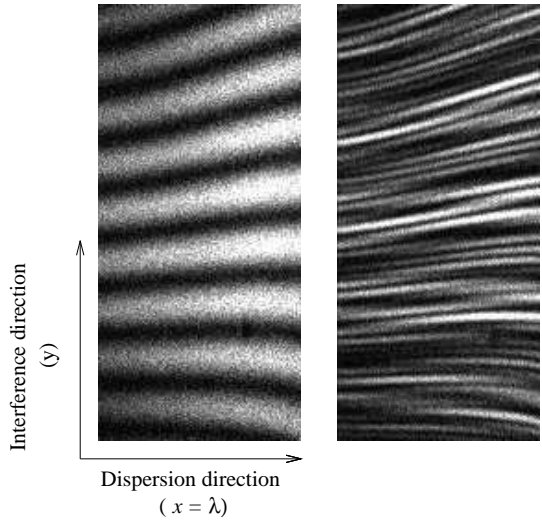

**Fig. 5.** Dispersed fringes obtained from a 50 μm pinhole illuminated by a tungsten filament. The exposures were made using masks with two holes (left) and five holes (right). The fringes diverge to the right with increasing wavelength and are curved because the dispersion of the prism is non-linear. The detector format was 110 × 240 pixels, giving a wavelength coverage of $\Delta\lambda = 120$ nm centred at $\lambda = 590$ nm. When observing, we normally use a narrower format (50×240), which provides a faster frame rate at the expense of a reduced wavelength coverage (see Sect. 2.5)

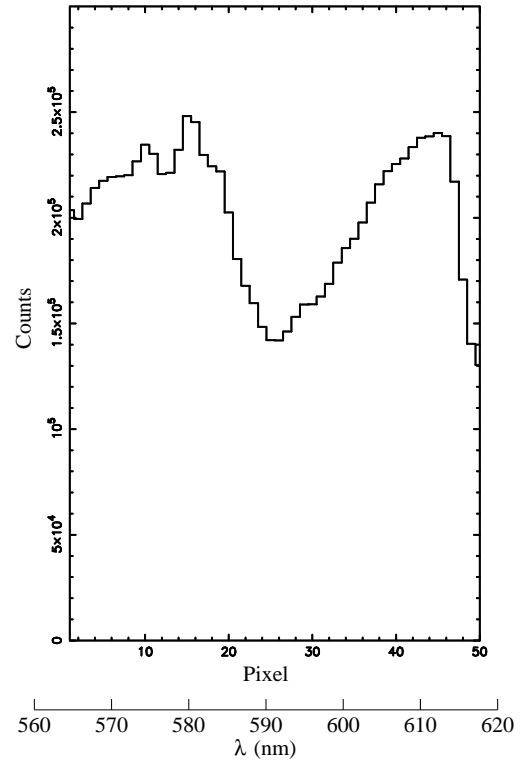

**Fig. 6.** Spectrum of $\beta$ Gru taken using MAPPIT, showing photon counts in each column of the IPCS and an approximate wavelength scale. The spectrum was obtained by combining summed images from 6 runs (total exposure time 1800 s). The wavelength resolution, as determined using emission lines from a copper-argon calibration lamp, is 2.5 pixels (FWHM).

The observing wavelength is selected using the second of the flat mirrors seen in Fig. 1, which has a micrometer tilt adjustment that allows one to direct the desired portion of the spectrum onto the detector. Figure 6 shows a spectrum of the M5 giant $\beta$ Gru (absorption features are due to TiO). The wavelength resolution $\delta\lambda$ and total wavelength coverage $\Delta\lambda$ of the system depend on several factors, including the observing wavelength and the size of the holes in the mask. Values for typical observations centred at 600 nm are $\Delta\lambda = 55$ nm and $\delta\lambda = 3$ nm. Thus, we achieve a wavelength resolution somewhat better than that of an interference filter. Moreover, the sensitivity is also improved because transmission losses are more severe for a filter than for the extra optics used in the dispersion system. Most importantly, we cover a large range of wavelengths simultaneously (the fractional bandwidth is $\sim 9\%$), and this is the main incentive for using wavelength dispersion.

It is important when aligning the system to ensure that the array of holes in the mask is perpendicular to the direction of dispersion. At a given wavelength, the images of the holes—as formed by the cylindrical lens—must overlap for interference to occur. We now see that overlap of the hole images will be more complete if we use square rather than circular holes. In practice, the alignment is done by adjusting the position angle of the mask in small increments and recording fringes from a pinhole source. The optimum mask angle can then be chosen by examining the power spectra of these fringe patterns.

From such alignment tests, we have determined that optimal overlap of the mask-hole images requires the holes to be slightly non-collinear. Computer ray tracing of the optics has confirmed that the holes should actually lie on a shallow arc. This is due to aberrations introduced by the prism, and indicates that the two orthogonal axes of the system cannot be treated as entirely independent. The necessary curvature, amounting to a maximum deviation from collinearity of $\sim 1°$, is included when we produce the artwork for the masks.

### 2.5. The IPCS detector

The detector used in MAPPIT is the Image Photon Counting System (IPCS; Boksenberg 1990). Light falls on the photocathode of a high-gain image intensifier, the output of which is optically coupled to a continuously scanning television camera. We operate the IPCS in its high-speed mode, in which the system records the coordinates of photon events together with markers indicating the start of each video frame. This information is written directly to magnetic tape.

The time resolution is set by the length of each video frame, which in turn depends on the number of pixels scanned. A detector size of 240×240 pixels gives a frame time of 16 ms, although the effective time resolution is somewhat worse than this because the video picture is scanned line by line, rather than being read out instantaneously. Using a smaller region of the detector allows faster sampling and most of our observations were made using an area of 50×240 pixels, which reduces the frame time to 6.5 ms. Of course, using this narrow format reduces the effective area of the detector. However, by adjusting the $x$-gain on the camera tube, we are able to increase the width of the pixels and so partially regain this lost area.

Several other characteristics of the IPCS are relevant for our experiment. On the positive side, the readout noise is zero and the dark count is very low (less than one photon per frame), which is vital when the number of photons per frame from the object itself may be only ten or so. On the negative side, the quantum efficiency of the image tube peaks around 400 nm, at a little over 20%, falling below 5% beyond 600 nm. This prevents us from taking advantage of the superior atmospheric conditions at longer wavelengths. More seriously, it reduces our sensitivity when studying the surfaces of cool stars.

Because it writes directly to tape, the data acquisition system is limited to $\sim 10^4$ photons per second (70 photons per frame). This limit is a problem when we observe bright objects, and we reduce the count rate by chopping the light beam with a rotating shutter. By sliding the shutter into the beam, we can vary the duty cycle continuously from 100% down to zero (see Fig. 1).

The speed of the chopping shutter is chosen to be about ten times slower than the IPCS frame rate, ensuring that all photon events are recorded during the times when the shutter is open. The average number of photons in each exposed frame can then reach $\overline{N} \approx 200$. Reducing the count rate in this way is much better than using a neutral density filter, for example, because the signal-to-noise ratio of the fringe power spectrum increases linearly with $\overline{N}$ and increases only as the square root of the number of frames. For closure phase measurements, the gain is even greater (signal-to-noise ratio $\propto \overline{N}^{3/2}$; see Haniff 1988).

We have recently installed a new acquisition system in which the data from the IPCS are intercepted by a Personal Computer and written directly to 8mm Exabyte tape via a SCSI interface. We are now able to record photon events at $\sim 50$ kHz, which is 330 photons per frame in the narrow (i.e., fast) format. By taking advantage of continuing improvements in available hardware, we would expect to be able to improve this count rate still further.

One minor problem is persistence in the phosphor of the IPCS television camera, which results in a single photon being registered over several consecutive frames. Recognizing and discarding the spurious photons is straightforward, however, and is done routinely as the first stage of data processing. Depending on the threshold levels in the IPCS electronics, the fraction of photon events that must be discarded can reach 50%.

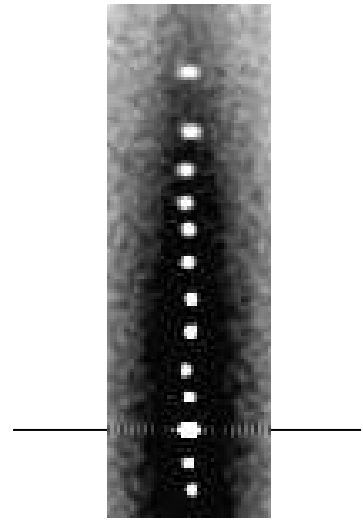

**Fig. 7.** Power spectrum of fringes from a 400 s observation of the unresolved star $\alpha$ Gru. The power spectrum is symmetrical about the origin (indicated by horizontal line); most of the lower portion is not shown here.

Even then the de-persistence software is unlikely to discard a significant number of genuine photon events, because most pixels in each frame are empty.

## 3. Data processing

Processing of non-redundant mask data involves extracting fringe visibilities and closure phases from the power spectrum and bispectrum, respectively, and then fitting them to a model of the object or reconstructing an image (Haniff et al. 1987; Nakajima et al. 1989). The main difference with MAPPIT data is the use of wavelength dispersion. The observations span a wide range of wavelengths and so the fringes diverge significantly across the detector. The simplest way to deal with this is to linearize the fringes before calculating the power spectrum or closure phases.

Fringe linearization is performed in two steps. First, we must establish the wavelength scale on the detector, which is done using spectra from a copper-argon arc lamp and dispersed fringes from an artificial star. Once we have calibrated a relation between the pixel column ($x$ coordinate) and wavelength, it is then a matter of reassigning the $y$ coordinate of each photon according to its wavelength. That is, each photon is shifted vertically by an amount that depends on its position on the detector. When this process is applied to a fringe pattern like that in Fig. 5, the result is series of parallel fringes (see Bedding 1992 for further details). Note that this method of processing the interference pattern involves several assumptions about the object being observed, as we discuss in Sect. 5.

Figure 7 shows the power spectrum of data from an unresolved star, calculated after making the correction for wavelength dependence. Note the slight negative bowl in the power spectrum, which arises because of coincidence losses in the de-tector. The event-centering electronics in the IPCS is designed to ensure that the 'splash' of electrons from each incident photon is recorded at a single address and, in doing so, renders the detector insensitive to closely separated photons. This is taken into account by fitting a slowly-varying function to the base level before extracting the visibilities. Thanks to the use of an aperture mask, this procedure is very easy because the signal in the power spectrum is confined to isolated regions. This is not the case for un-masked (i.e., speckle) observations, and the correction for detector bias is then more difficult (Pehlemann et al. 1992).

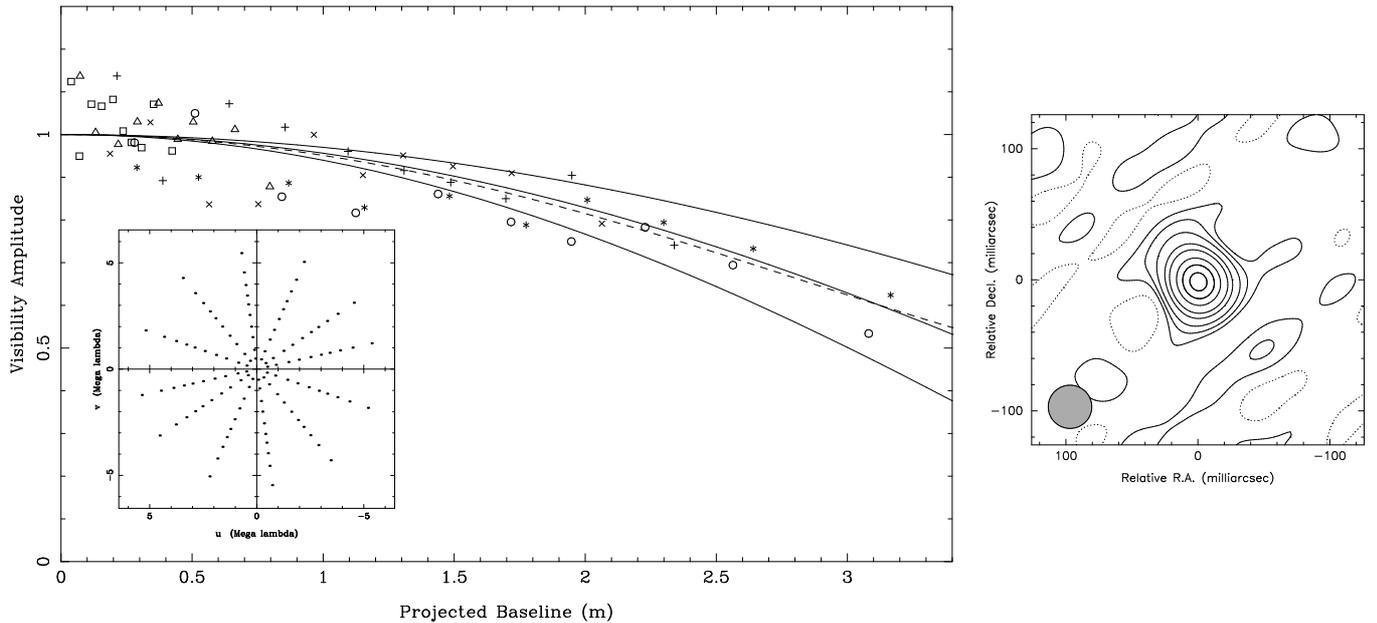

**Fig. 8.** Observations of $\sigma$ Sgr (central wavelength 563 nm, $\Delta\lambda = 53$ nm). The left panel shows calibrated visibility amplitudes projected along the position angle 22°. A different symbol is used for each data set (error bars are not shown, but are similar to those in Fig. 9). Solid curves indicate the amplitude expected from a double star with equal components and a separation of $(11.5 \pm 2)$ mas. The dashed curve is for an unequal double with $\Delta m = 1$ and separation 13.5 mas. In the reconstructed image (right), north is down and east is to the right; the shaded patch shows the restoring beam (FWHM 33 mas). The $(u, v)$ coverage of the observations is shown in the inset.

Another feature of the power spectrum in Fig. 7 is that the ten peaks are not quite collinear. This has nothing to do with the mask curvature described in Sect. 2.4. Rather, it is due to dispersion in the optics; the light from each hole in the mask passes through a different thickness of glass, and so is delayed by an amount which depends on wavelength. For each baseline, this delay manifests itself as a rotation of the corresponding set of fringes. Each baseline is affected differently, resulting in the slight 'zig-zag' pattern in the power spectrum. It is straightforward to allow for this effect when determining the visibility amplitudes. Note that this dispersion is the cause of the fringe crossing seen in Fig. 5.

## 4. Results

MAPPIT observations of the following double stars have been published elsewhere: $\eta$ Oph (Robertson et al. 1991), $\delta$ Sco (Bedding 1993) and $\iota^1$ Lib (Bedding et al. 1994). Here, we report on measurements of another double star ($\sigma$ Sgr) and on two resolved single stars ($\alpha$ Sco and $\beta$ Gru).

### 4.1. The double star $\sigma$ Sgr

This star was found to be double by Hanbury Brown et al. (1974) using the Narrabri intensity interferometer. They reported the star to be a binary with components of roughly equal brightness ($\Delta m = 0.6 \pm 0.5$), although their measurement gave no information about the geometry of the components. Although the star is bright ($V = 2.8$), we can find no mention in the literature of other observations.

We observed $\sigma$ Sgr with MAPPIT on 1991 July 26 with seeing of $\sim 2$ arcsec. As a reference star we used $\lambda$ Sgr, which has an angular diameter of 4.4 mas[1] (Harwood et al. 1975) and so is effectively point-like for MAPPIT. The visibilities of $\sigma$ Sgr on the longest baselines were substantially reduced at some position angles, consistent with the object being a barely-resolved double. In the following analysis, we assume this to be the case. Note that a star having the spectral type and magnitude of $\sigma$ Sgr (B2.5 V and $V = 2.0$) is expected to have an angular diameter of less than 1 mas, which is completely unresolvable by MAPPIT.

Model fitting and image reconstruction were done using the Caltech VLBI software. Because the double is not fully resolved, it is difficult to determine the magnitude difference accurately. We therefore began by assuming $\Delta m = 0$, consistent with the Hanbury Brown et al. (1974) result, and fitting only to the visibility amplitudes. From this two-dimensional model we determined a position angle of $(22\pm10)°$ and separation $(11.5\pm2)$ mas. A convenient way to display this fit is by projecting the data along the best-fit position angle (22°), as shown by the solid curves in Fig. 8.

---

[1] We use "mas" as an abbreviation for milliarcsecond throughout

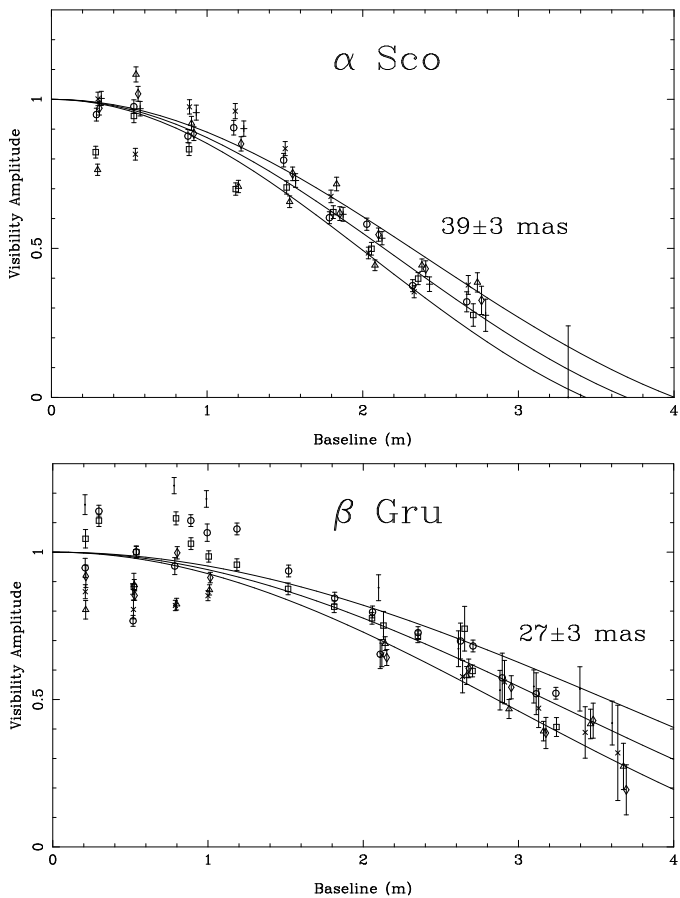

**Fig. 9.** Calibrated visibility amplitudes for two resolved stars, fitted by uniform disks. The observations were made in wavelength bands 565–582 nm ($\alpha$ Sco) and 565–618 nm ($\beta$ Gru). Different symbols represent observations at different position angles and the data points have been spread horizontally by a small amount to improve readability. No fringes were detected from $\alpha$ Sco on the longest baseline (3.3 m); the error bar shows an upper limit.

Note that non-zero values of $\Delta m$ fit the observations equally well. For example, the dashed curve in Fig. 8 shows the visibilities expected with $\Delta m = 1.0$ and separation 13.5 mas. If the double is unequal, visibility amplitudes alone cannot resolve the 180° ambiguity (i.e., establish the parity of the binary). We therefore repeated the modelling process with the closure phases included. The best fit to an unequal double was for $\Delta m = 0.2$ and position angle 202° (i.e., with fainter component to the south). This orientation is visible in the asymmetry in the image shown in Fig. 8, which was made using conventional hybrid mapping techniques (CLEAN and self-calibration; see Pearson & Readhead 1984). However, it must be noted that the signal-to-noise ratio in the closure phases is not high enough to determine $\Delta m$ accurately or to establish the parity with certainty.

### 4.2. Single stars: $\alpha$ Sco and $\beta$ Gru

We observed the M1.5 supergiant $\alpha$ Sco on 1991 June 1 at three different position angles, using observations of $\varepsilon$ Sco to calibrate each measurement. The seeing was 1–1.5 arcsec. We find no evidence in our data for variation in the visibility of $\alpha$ Sco with position angle. Nor did we find the closure phases to differ from zero, within errors of $\sim 10°$.

Figure 9 shows the visibility amplitudes of $\alpha$ Sco divided by those of $\varepsilon$ Sco. Fitting these points to a uniform disk gives an angular diameter of $(39 \pm 3)$ mas. Note, however, that we expect the reference star $\varepsilon$ Sco also to be slightly resolved. This star has no published angular diameter, but we can use its spectral type (K2.5 III) and magnitude ($V = 2.3$), together with the calculations listed by Ochsenbein & Halbwachs (1982) to estimate a diameter of 9 mas. An earlier estimate by Wesselink et al. (1972) gives 7.3 mas. Correcting the visibility curve of $\alpha$ Sco for the non-zero size of the reference star leads to a slightly revised diameter of $(40 \pm 3)$ mas. This value is in good agreement with a recent lunar occultation measurement by Richichi & Lisi (1990), who found a uniform-disk diameter of $(38.9 \pm 1)$ mas at a wavelength of 2.43 $\mu$m.

The star $\beta$ Gru has spectral type M5 III and magnitude $V = 2.1$. We observed this star on two nights (1991 July 26 and 29) at a total of seven position angles under poor seeing conditions (2–3 arcsec). The reference star was $\alpha$ Gru (spectral type B7 IV, magnitude $V = 1.7$), which has an angular diameter of 1.0 mas (Hanbury Brown et al. 1974) and so is point-like for MAPPIT. As shown in Fig. 9, we find $\beta$ Gru to be resolved with a uniform-disk diameter of $(27\pm3)$ mas. To our knowledge, this is the first time that this star has been resolved.

As can be seen from Figs. 8 and 9, there is a large scatter in the visibility measurements for the shortest baselines. This arises from a time-varying transfer of power between the shortest baselines and results in significantly greater errors than expected purely from photon and atmospheric noise. The cause of this effect is unknown, but fortunately it does not greatly affect the accuracy on the more important longer baselines.

## 5. Discussion

The method for processing dispersed fringes described in Sect. 3 involves averaging together all wavelength channels. This requires making several assumptions about the object being observed. The first point is that, in processing all wavelength channels as one, we are averaging the object visibility over a range of spatial frequencies. If the visibility of the object changes significantly over this spatial frequency range, our estimate of the visibility will be in error. This is known in radio astronomy as bandwidth smearing because it introduces radial distortions at the edges of the reconstructed image (Bridle & Schwab 1988).

The severity of bandwidth smearing depends on the angular extent $\theta$ of the object and the angular resolution $\lambda/d$ of the measurement. Here, $d$ is the maximum baseline used for the observation. It is not hard to show that the errors in the estimated visibilities will be small provided

$$\frac{\Delta\lambda}{\lambda} \ll \frac{\lambda}{d}\frac{1}{\theta}.$$

In MAPPIT, this means that the field of view is typically limited to $\theta \approx 0.3''$.

The second assumption inherent in combining all wavelength channels is that the structure of the object does not change with wavelength. Such an assumption is not valid if, for example, we wish to measure the wavelength dependence of the angular size of a star. Quirrenbach et al. (1993) have found cool giants (spectral types M3 III to M5 III) to appear $\sim 10\%$ larger in TiO absorption bands relative to the continuum. The simplest way to extract information on wavelength dependence from MAPPIT's dispersed fringes is to subdivide the detector into several wavelength intervals and process each interval separately. We used this method with the data from $\beta$ Gru (spectral type M5 III) and found no difference in angular diameter between the continuum and TiO bands. However, the accuracy of this measurement was only 10–20% and was limited by the signal-to-noise in the fringe visibilities (seeing conditions were poor). In the case of $\alpha$ Sco, the observed wavelength band was narrower and did not contain any strong TiO absorption features, so no similar analysis was made.

For high-quality data, it should be possible to extract information about wavelength-dependent structure. We are currently investigating new techniques in an attempt to use all the information contained in the dispersed fringes. The most promising scheme involves iterative model fitting, in which an initial model is compared with the fringe pattern and the mismatches are used to modify the model and produce a closer fit on the next iteration. This scheme is analogous to the self-calibration algorithm that is widely used in radio astronomy (Pearson & Readhead 1984).

## 6. Conclusion

Aperture masks are an important tool for high-resolution observations, at least for some types of observations. The results presented here are for bright objects that are only just resolved by the telescope aperture. For these objects, the mask allowed us to achieve very good angular resolution. In this regard, using a fully-filled aperture would have two disadvantages: (i) it gives diminished weighting to long baselines relative to shorter ones, and (ii) it is more difficult to calibrate the visibility amplitudes because of atmospheric noise caused by the redundancy of the pupil.

The result for $\sigma$ Sgr reported here represents the smallest separation measured for a binary star with a single telescope (McAlister & Hartkopf 1988). It provides a good demonstration of how aperture masking allows high resolution observations of an object that is only just resolved. However, it must be noted that aperture masking is only useful for a limited set of binary stars. Very close doubles require a long-baseline interferometer (e.g., Armstrong et al. 1992). For wider doubles, full-aperture speckle has only given useful results for separations above $\sim 25$ mas (see McAlister & Hartkopf 1988), but in this regime it is more efficient than aperture masking because it provides complete instantaneous $(u, v)$ coverage.

Aperture masking has so far been limited to simple bright objects (double stars and resolved single stars), where it has been very successful (e.g., Wilson et al. 1992). The technique of wavelength dispersion, as described here, and the use of a thin slit instead of holes (Buscher & Haniff 1993) both offer the possibility of extending aperture masking to fainter objects.

*Acknowledgements.* MAPPIT development was supported by a grant under the CSIRO Collaborative Program in Information Technology. The project was initiated in collaboration with Bob Frater, John O'Sullivan and Ray Norris and we thank them for their continued interest and support. The idea of combining image-plane and pupil-plane systems came from a suggestion by Yves Rabbia. We are grateful for invaluable assistance provided by Peter Gillingham and other AAT staff and by John Horne, and to the Australian Time Assignment Committee and the AAO Director for allocating observing time. We thank the referee for several important comments on this paper.